\documentclass[a4paper,11pt]{article}
\usepackage{pos}

\title{The Pacific Ocean Neutrino Experiment}
 \ShortTitle{P-ONE}

\author*[a]{Elisa Resconi}
\forColl{P-ONE}

\affiliation[a]{Technical University Munich, Physics Department \\
Excellence Cluster ORIGINS  \\
  James-Franck-Str. 1, Garching bei München, Germany}

\emailAdd{elisa.resconi@tum.de}

\abstract{Neutrino telescopes are unrivaled tools to explore the Universe at its most extreme. The current generation of telescopes has shown that very high energy neutrinos are produced in the cosmos, even with hints of their possible origin, and that these neutrinos can be used to probe our understanding of particle physics at otherwise inaccessible regimes. The fluxes, however, are low, which means newer, larger telescopes are needed. Here we present the Pacific Ocean Neutrino Experiment, a proposal to build a multi-cubic-kilometer neutrino telescope off the coast of Canada. The idea builds on the experience accumulated by previous sea-water missions, and the technical expertise of Ocean Networks Canada that would facilitate deploying such a large infrastructure. The design and physics potential of the first stage and a full-scale P-ONE are discussed.}

\FullConference{37$^{\rm{th}}$ International Cosmic Ray Conference (ICRC 2021)\\
		July 12th -- 23rd, 2021\\
		Online -- Berlin, Germany}

\begin{document}
\maketitle
\noindent
The intrinsic properties of 1) the high-energy universe beyond our Galaxy and 2) obscured astronomical objects can not be fully revealed using photon-based astronomy only. 
Neutrinos interact only weakly; hence are ideal astrophysical messengers but higher statistics is needed to access to the nature of active galaxies, blazars, dark matter and explore the cosmos up to the highest energy frontier.
The observatory leading high energy neutrino astronomy research is the IceCube Neutrino Observatory (IceCube) at the South Pole (see \url{https://icecube.wisc.edu}). 
The milestones achieved with the neutrinos collected by IceCube during a period of more than 10 years of data searching for cosmic accelerators are:
\begin{itemize}
    \item The discovery of a diffuse astrophysical flux of neutrinos \cite{Aartsen:2013bka, Aartsen:2013jdh, IceCube:2020acn, IceCube:2020wum, IceCube:2021jmr};
    \item The hint of the association of neutrinos to blazars \cite{Padovani:2014bha,  Padovani:2015mba, Padovani:2016wwn, Rodrigues:2020fbu, Giommi:2020hbx, Giommi:2020viy, Petropoulou:2020pqh};
    \item The evidence of the association of an alert event and a neutrino flare to the $\gamma$-ray emitting blazar TXS\,0506+056 \cite{IceCube:2018cha, IceCube:2018dnn, Padovani:2018acg};
    \item The hint of the association of neutrinos to the Seyfert galaxy NGC\,1068 \cite{IceCube:2019cia, Inoue:2019yfs}.
\end{itemize}
While this progress is exciting, it also reveals that with a telescope only of the size of the order of a cubic kilometer, the neutrino sample collected from the cosmos is still too limited to advance on this promising, rich path of fundamental discoveries in astro and particle physics. 
To gain more statistics and exposure to the entire sky, a global effort is mobilizing to establish dramatic improvements in the integral exposure to astrophysical neutrinos. New feasible technologies, methodologies, and observatories, ultimately linked through a global network are now to be fostered \cite{Schumacher:2021Pv}. 
\begin{figure}[b!]
\centering
\includegraphics[width=0.6\textwidth]{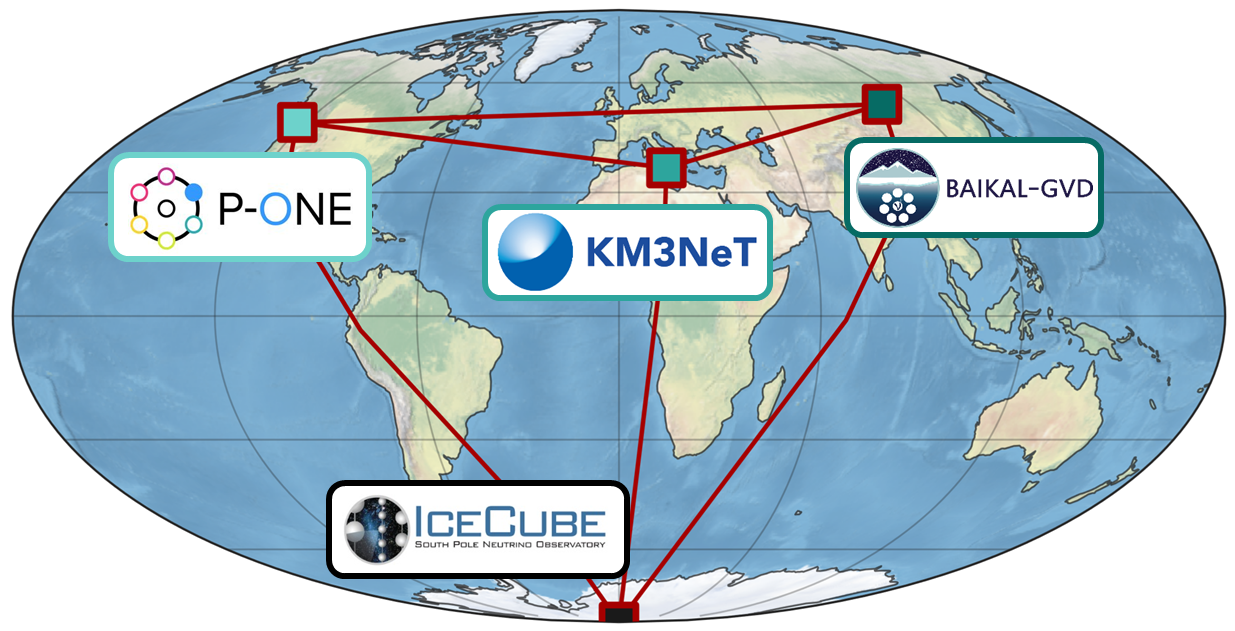}
\caption{{\small{\bf Map of existing or under construction neutrino telescopes \cite{Schumacher:2021Pv}.  }}
\label{Fig:map}}
\end{figure}
A network of telescopes, as the one in Fig.\,\ref{Fig:map}, will expand neutrino observation beyond the limits of the northern sky covered by IceCube to include also the southern sky. \\
Neutrino astronomers and oceanographers alike have envisioned exploiting deep sea infrastructure to support environmental and fundamental research. Thousands of kilometers of underwater telecommunications cables are seen as a potentially ideal host for massive, cubic-km scale neutrino telescopes and as an interface to study the secrets of ocean systems and global environmental systems,  however the neutrino astronomy community has missed the infrastructure available to be used as a piggyback. We therefore conceived the Pacific Ocean Neutrino Experiment (P-ONE) as the first neutrino telescope hosted by an existing oceanographic infrastructure.

\section{Visualize P-ONE}
\begin{figure}[htp]
    \centering
    \includegraphics[width=0.83\textwidth]{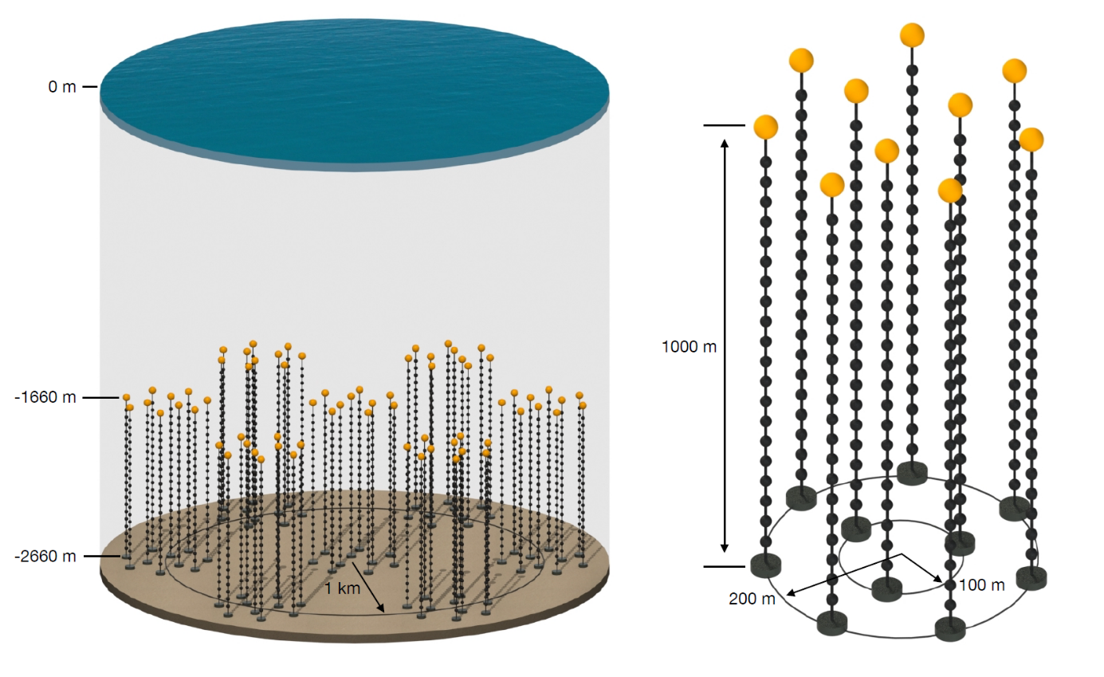}
    \caption[P-ONE cluster concept]{{\small{\bf P-ONE first vision \cite{P-ONE:2020ljt}. The left side shows the complete telescope configuration, consisting of 7 clusters. An individual cluster with 10 mooring lines is depicted on the right. A full optimization of the telescope is presently on-going. }}}
    \label{fig:cluster}
\end{figure}

\noindent
The Pacific Ocean Neutrino Experiment (P-ONE) \cite{P-ONE:2020ljt} is a proposed multi-cubic kilometer-scale neutrino telescope to be deployed off the coast of Vancouver Island, Canada (see Sec.\,2).  The main goal of P-ONE is to significantly advance the field of neutrino astronomy by extending the cosmic frontier at the highest energies. P-ONE is a discovery-oriented mission aiming to reveal previously unknown astronomical phenomena, test fundamental physics at the PeV scale, and provide crucial information for multi-messenger follow-up observations. P-ONE leverages the expertise in deep-sea operations from Ocean Networks Canada (ONC). P-ONE will complement the existing network of neutrino telescopes around the globe (Antares \cite{ANTARES}, KM3NeT \cite{KM3NET}, GVD Baikal \cite{GVD}, IceCube \cite{ICECUBE}) and significantly contribute to a complete neutrino sky coverage at the PeV scale \cite{Schumacher:2021Pv}.
P-ONE is currently envisioned as a segmented structure, consisting of 7 clusters with ten mooring lines each. Each line will include 20 optical modules of two different types: a calibration module and an optical receiver module, see Fig.\,\ref{fig:cluster}. 
P-ONE is designed to detect astrophysical neutrinos above the TeV energies with excellent directional and energy reconstruction for neutrino events, focusing on horizontal tracks where discovery potential is maximal. Another critical aspect of the design is a modular structure that allows for installation of varying sizes and easy scalability of the array in different stages. A single P-ONE cluster will consist of 10 strings with 20 optical modules each. Two pathfinder experiments are currently studying the optical properties of the Cascadia Basin, Strings for Absorption Length in Water (STRAW) and STRAW-b, deployed in 2018 and 2020 (see Sec.\,3) \cite{Bailly:2021Qi,Rea:2021o3}. The next milestone towards P-ONE is the development of the prototype instrumented line and the construction of at least three for in-situ validation of performances (see Sec.\,4) \cite{Spannfellner:2021ro}. The role of P-ONE in the global network of neutrino telescopes has been addressed by the PLE$\nu$M team, summarized in Sec.\,5 \cite{Schumacher:2021Pv}.

\section{Ocean Networks Canada, NEPTUNE, and the Cascadia Basin Node}
\noindent
{\bf Ocean Networks Canada (ONC)}
Ocean Networks Canada (ONC)\footnote{\url{http://oceannetworks.ca/}} is a multidisciplinary organization with an exemplary track record in deploying large-scale, real-time sensor networks in the deep ocean with increasing reliability over the last decade. The ONC has a success rate of off-shore maintenance cruises of the order of 95\%. Those rates are the result of a set of strictly applied standard operating procedures that put the emphasis on thorough pre-testing, and are supported by rigorously observed work-flows.
ONC is a leading global organization supporting research and discovery in the oceans, including, since 2017, the P-ONE mission.
It is composed of many observatories, the largest being the North East Pacific Time-series Underwater Networked Experiment (NEPTUNE)\cite{Neptune:2010}.\\

\begin{figure}[htp]
\centering
\includegraphics[width=0.8\textwidth]{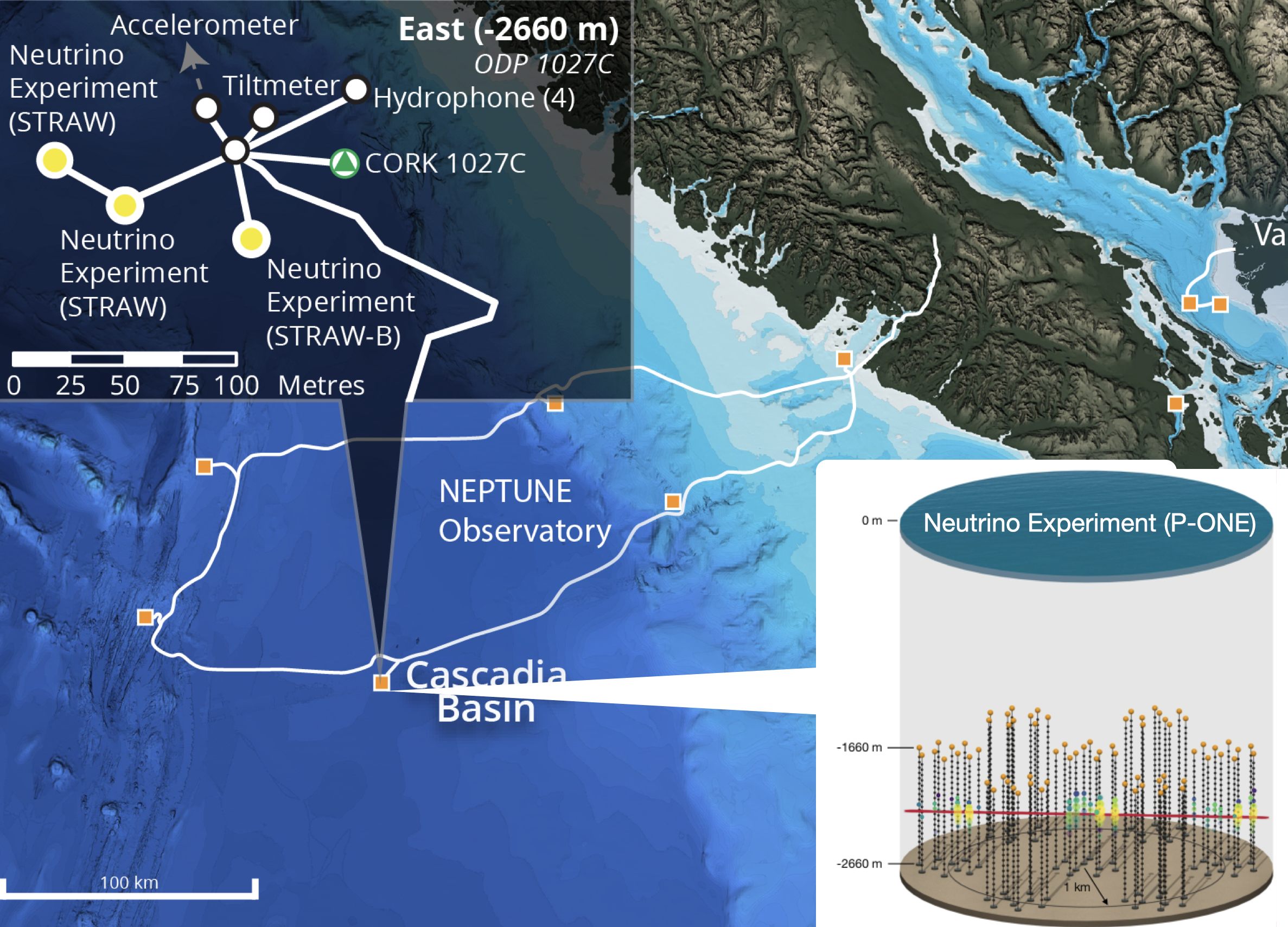}
\caption{{\small{\bf Map of the 800 km long NEPTUNE cable observatory operated by Ocean Networks Canada (white loop) with an exploded view of the instrumentation presently taking date at Cascadia Basin (black window) and a vision of P-ONE telescope (white window). }}
\label{Fig:NEPTUNE}}
\end{figure}

\noindent
{\bf The NEPTUNE observatory} (see Fig.\,\ref{Fig:NEPTUNE}) is operated by ONC since 2009. It is comprised of an 800\,km loop of telecommunication fibre-optic cables to power and transfer data to a wide variety of sensors. The high-speed data link (up to 4 Gbit/s) and high power (of the order of 8 kW/node) supports 5 nodes acting as a local hub for core observations and experiments. A total of 17 primary junction boxes are cabled to the nodes and used to connect hundreds of instruments. The connections rely on underwater-mateable connectors with field proven reliability (a failure lower than 2\% in deployed connector pairs over a period of 10 years). 
The NEPTUNE observatory offers an ideal and available oceanographic infrastructure and in this way the opportunity for the construction of a large volume neutrino telescope. 
In 2017, we set up a new collaboration with ONC for its use in the context of neutrino astronomy. 
Among the various ONC-powered nodes, the Cascadia Basin at a depth of 2660 meters has been selected, after two years of in-situ measurements, to host the Pacific Ocean Neutrino Experiment (P-ONE, see \url{http://www.pacific-neutrino.org}). \\

\noindent
{\bf Cascadia Basin} is a deep area lying between the Juan de Fuca Ridge and the Pacific Northwest, which hosts one of the five instrumented nodes of NEPTUNE. Already cabled and constantly powered by ONC,  Cascadia Basin is a geological and environmental body of water known to be a calm sedimental area with weak currents (3-7 cm/s) and temperature around 2$^{\circ}$C \cite{https://doi.org/10.1029/2011GC003922}. It has been instrumented with several monitoring systems, such as hydrophones, tiltmeters, thermometers, accelerometers, and acoustic Doppler current profilers (ADCP), providing an excellent starting point for the assessment of the environmental conditions. 

\section{Find the path to P-ONE}
\noindent
Fascinated by the ONC success rate, their world-wide uniqueness and their ease of operation deep in the ocean, a collaboration between TU Munich, University of Alberta, Simon Fraser University, Queen's University, and ONC was formed in late 2017. 
If a lot is already measured and understood about the environment in Cascadia Basin from ONC, the determination of the optical properties has required a dedicated effort. For this reason, the P-ONE collaboration has launched in 2018 a first pathfinder mission, the {\it STRings for Absorption length in Water} (STRAW), and in 2020 a second one (STRAW-b)  \cite{Bedard:2018zml, Rea:2021o3}. Both missions resulted in installation of optical and calibration sensors. 

\begin{figure}[htp]
\centering
\includegraphics[width=0.5\textwidth]{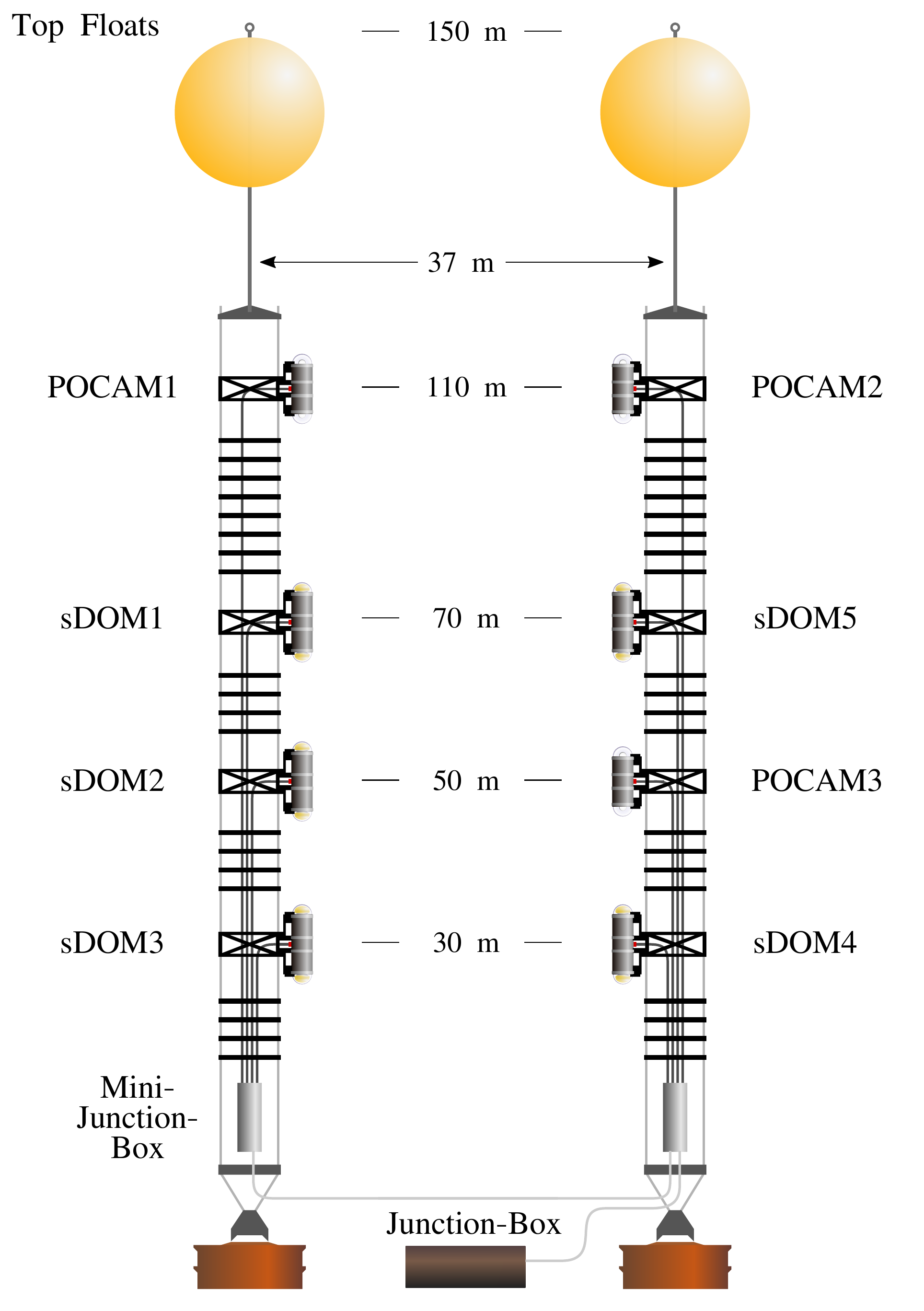}
\caption{{\small{\bf The two STRAW mooring lines equipped with three calibration modules (POCAM) and five photosensors (sDOM). The lines have been daisy chained and connected to a mini Junction Box (mJB). The mJB is connected to a primary JB already integrated in the ONC infrastructure.  The geometry was optimized to obtain various baselines between the light emitter (POCAM) and receiver (sDOM). The lines were oriented face-on at the time of the submarine connection, thanks to a rotatable anchor.   }}
\label{Fig:STRAW}}
\end{figure}

\noindent
{\bf The STRAW Pathfinder}\\
As a first pathfinder mission for P-ONE, ONC top-down deployed two 145\,m tall instrumented mooring lines in Spring 2018 , see Fig.\,\ref{Fig:STRAW} \cite{Bedard:2018zml}.  The lines have been designed, commissioned by the nascent P-ONE collaboration. Thanks to strict quality control based on standardized testing and final qualification tests in ONC's external facilities, all instruments, three Precision Optical CAlibration Modules (POCAMs), and five photosensors are reliably operating for more than three years with a duty cycle of 98\%.
Data are flowing from the deep Pacific Ocean to the P-ONE collaboration, and we achieved the preliminary assessment of the attenuation length vs. wavelength, rate of bioluminescence, and $^{40}K$ \cite{Authors:2021htl, Bailly:2021Qi}.\\

\begin{figure}[htp]
\centering
\includegraphics[width=0.42\textwidth]{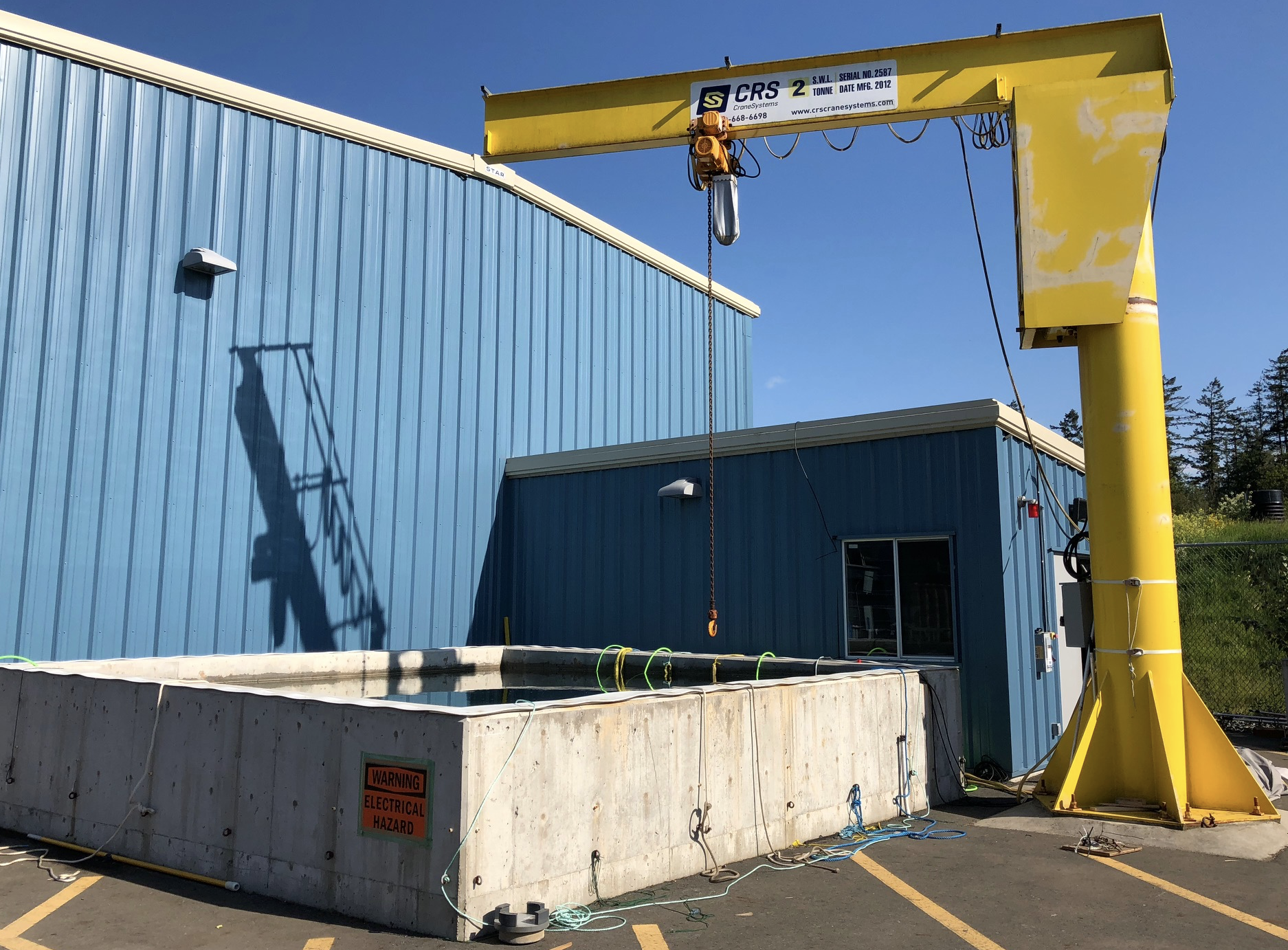}
\includegraphics[width=0.465\textwidth]{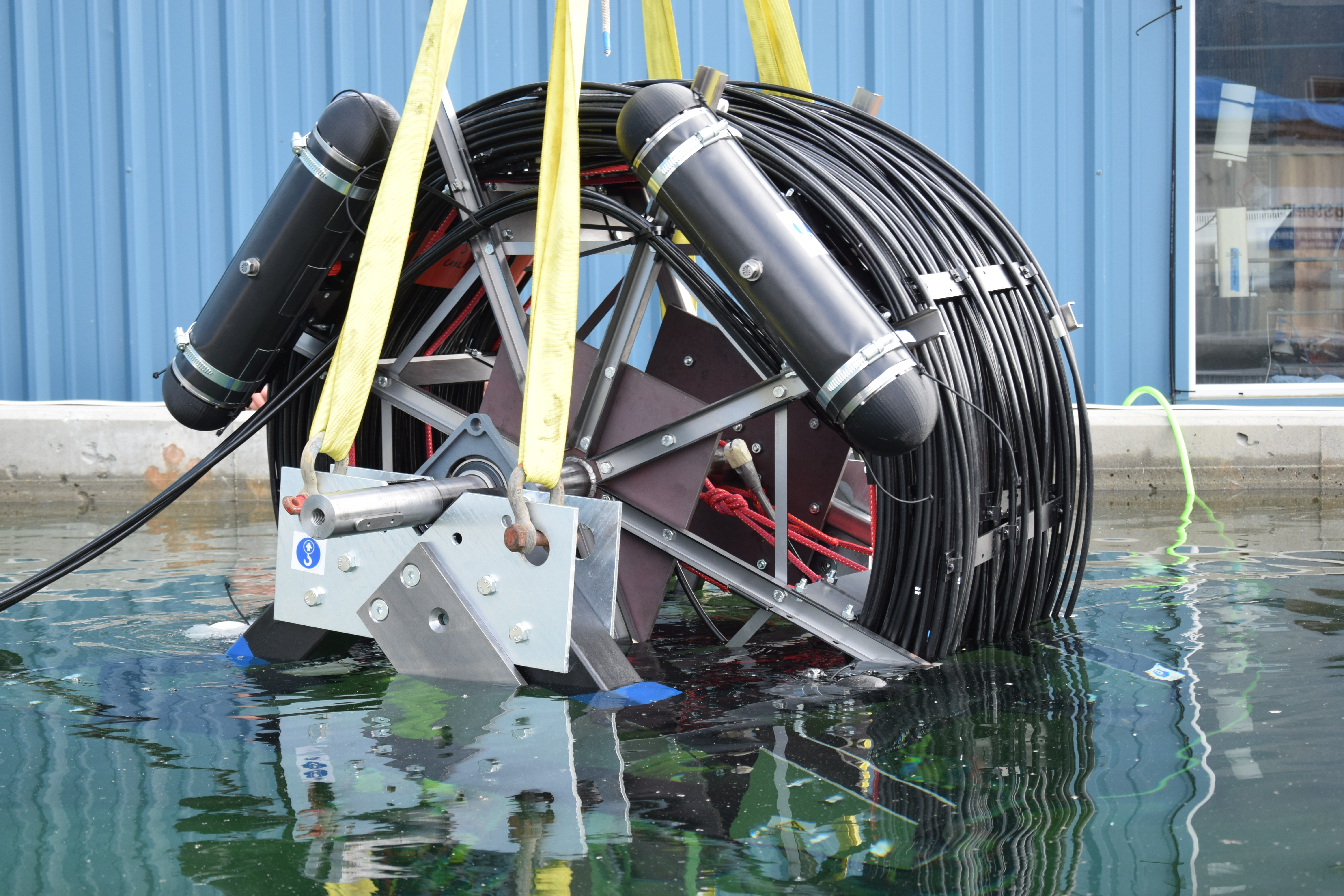}
\caption{{\small{\bf Quality control at the Marine Technology Center (MTC) of ONC. On the left is the saltwater pool with an overhead crane. The crane is permanently mounted to lift and immerse heavy infrastructure, including fully instrumented mooring lines. On the right, one of the STRAW lines enters the pool for long-term tests. The submerged instrumentation can be powered and monitored in its operation during submersion tests. The pool can be covered to avoid damaging the photosensors.   }}
\label{Fig:quality}}
\end{figure}

\begin{figure}[htp]
\centering
\includegraphics[width=0.6\textwidth]{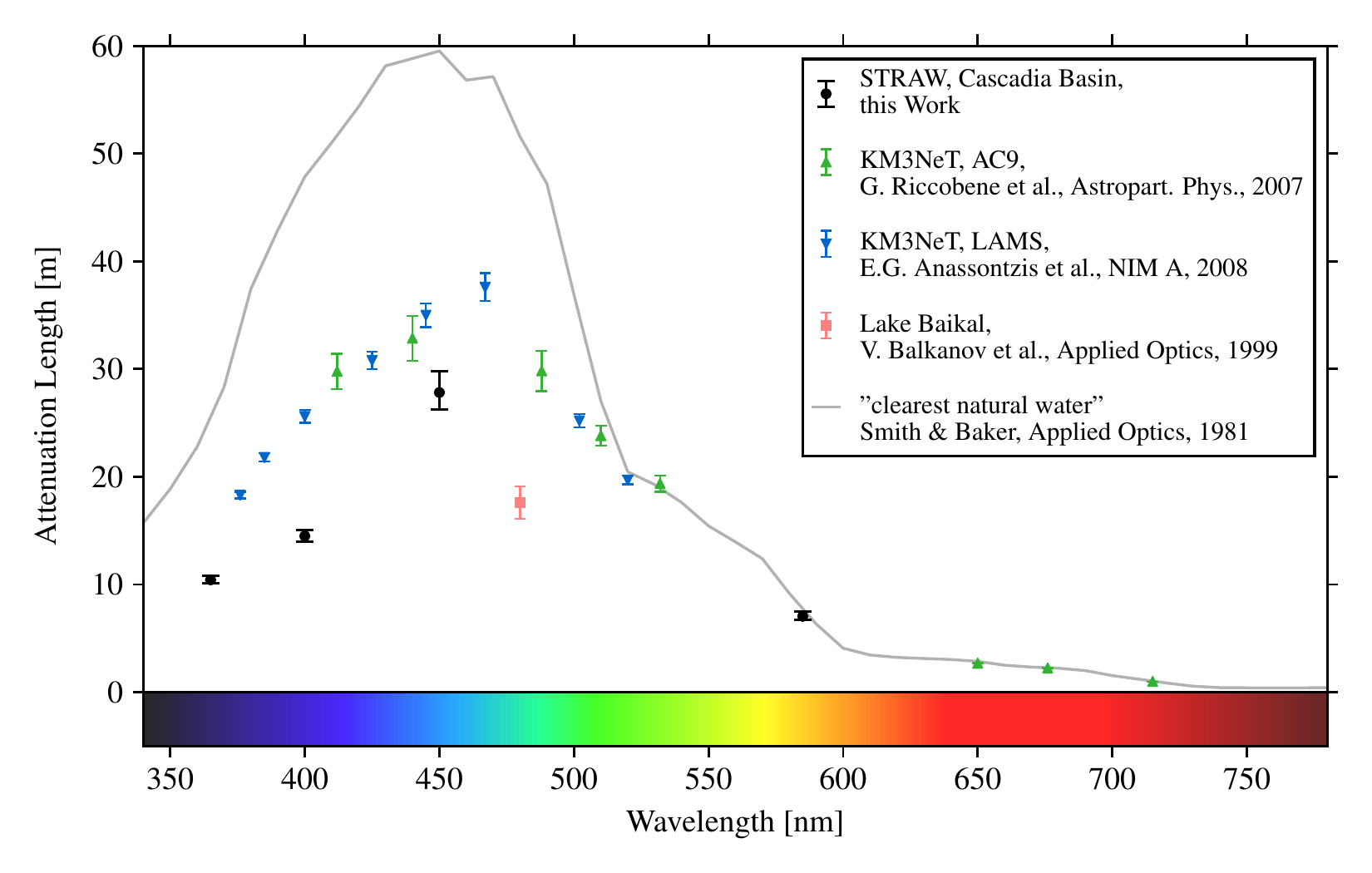}
\caption{{\small{\bf Attenuation length estimated from the STRAW data collected over four different wavelength and comparisons with other sites \cite{Authors:2021htl, Bailly:2021Qi} (see text for references and details).}}
\label{Fig:attenuation}}
\end{figure}

\noindent
Results from the STRAW pathfinders are reported in \cite{Bailly:2021dxn}. Most notably, we preliminary report the attenuation length measured at four wavelengths: $\lambda_{365 nm}\approx$ 11\,m, $\lambda_{400 nm}\approx$ 15\,m, $\lambda_{450 nm}\approx$ 30\,m, $\lambda_{585 nm}\approx$ 8\,m. Results are shown in Fig.\,\ref{Fig:attenuation} where measurements from other sites are also show for comparison \cite{balkanov_situ_1999,riccobene_deep_2007,anassontzis_light_2011}. Details on how the analysis has been performed and about the comparisons are available in \cite{Bailly:2021dxn}. If the observed value of the attenuation length at 450\,nm is similar to other sites, the values at lower wavelengths appear reduced. A possible explanation could be related to the fact that STRAW is relatively close to the Cascadia Basin floor, where more significant sediment dissolution is likely. 
We anticipate that the attenuation length is dominated by absorption in the water and that scattering is subdominant. Direct verification of both the attenuation length in the uppermost part of the water column and the scattering length motivated the construction of a second pathfinder, STRAW-b.   \\

\begin{figure}[htp]
\centering
\includegraphics[width=0.49\textwidth]{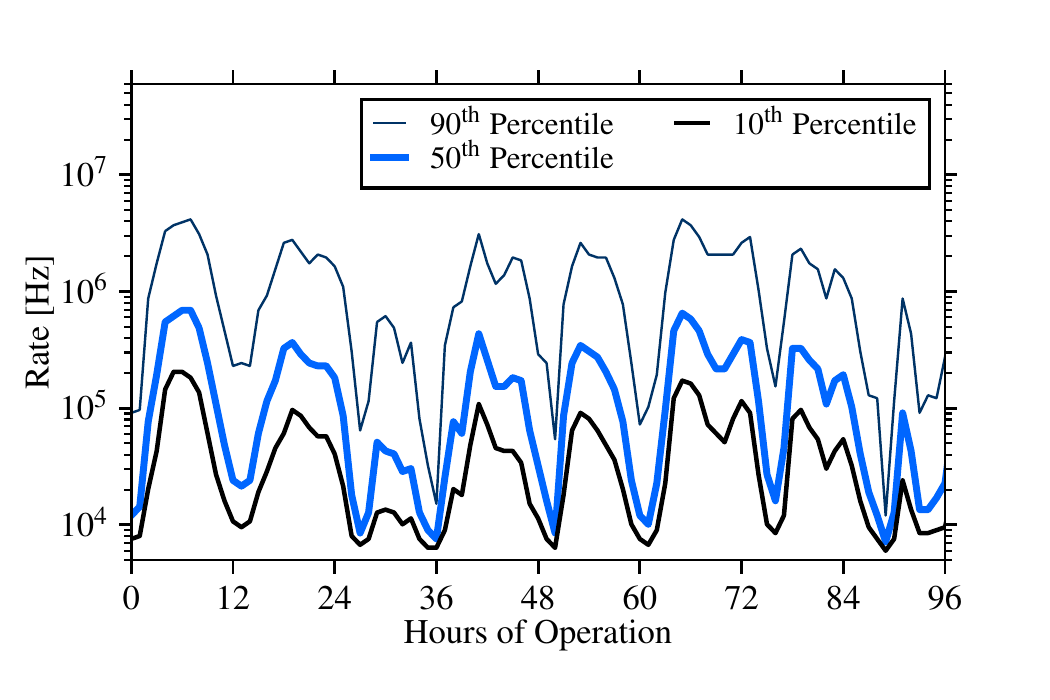}
\includegraphics[width=0.41\textwidth]{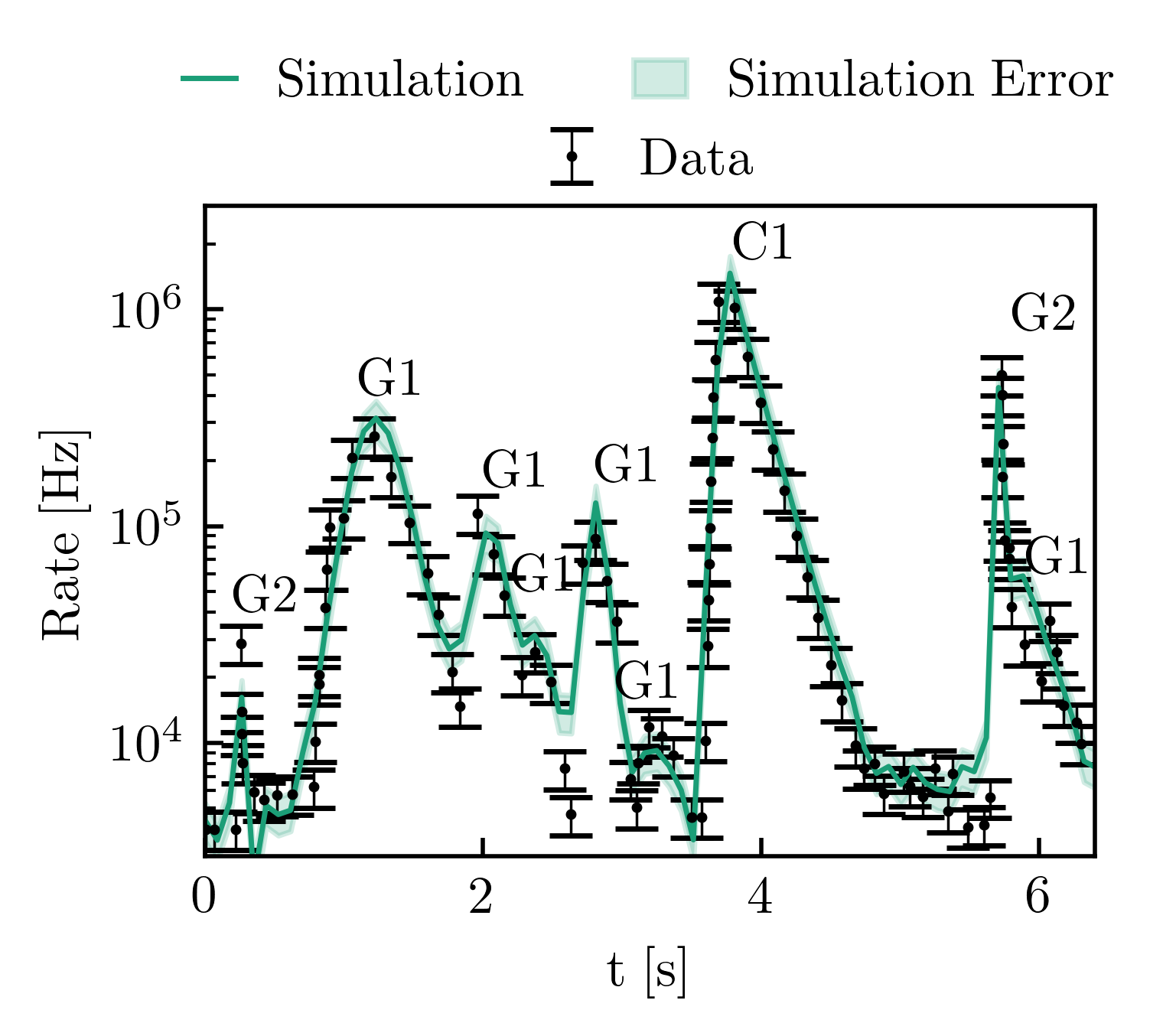}
\caption{{\small{\bf (left) Bioluminescence rate observed at Cascadia Basin, hourly percentiles over a representative four days of data taken. The time structure visible in the distribution corresponds to the 12.5 hour tidal cycle. We note that we report the percentiles instead of the mean rate since the percentiles are unaffected by the DAQ saturation limit \cite{Authors:2021htl, Bailly:2021Qi}. (right) 
A comparison of the data gathered by STRAW-b \citep{Rea:2021o3} and the simulation presented here. We performed a fit on data, while minimizing the number of organisms. This resulted in a population of three distinct species to model the data, denoted G1, G2, and C1." Is the textblob in the paper (the new version)
G1, G2 denote organisms which emit light in the form of a Gamma distribution (for their time-series)
C1 emits light in the form of a Cauchy distribution  \cite{meighen2021bioluminescence}.}}}
\label{fig:four-days}
\end{figure}

\noindent
Another environmental parameter of fundamental importance is the level of locally observed bioluminescence and the temporal variation of this point emission of light by organisms. Also covered in more details in \cite{Bailly:2021dxn}, we report in Fig.\,\ref{fig:four-days} (left), the hourly percentiles of the background rates over a period of four days. Such behaviour is representative of the observed behaviour recorded constantly (98\% duty cycle).
In parallel to the data taken and bioluminescence empirical study, members of the P-ONE collaboration have anticipated a modelling work predicting deep sea bioluminescence simulating the water flow around the mooring lines of the P-ONE telescope \cite{meighen2021bioluminescence}. 
Two types of causes producing flashes have been investigated: the encounter and the shear stress generated by the tides and the telescope's structure. 
Preliminary findings indicate that shear stress play a far larger role at the low densities expected in the deep sea. Moreover, investigation of unique bioluminescence signatures once compared with the simulation might reveal insights in the species producing the bioluminescence, topic of interest for the marine biologists in P-ONE. More studies related to the spectrum of the bioluminescence are on-going in STRAW-b pathfinder, see Fig.\,\ref{fig:four-days} (right). \\

\noindent
While the primary background consists of stochastic bioluminescence spikes, it is also expected that a continuous noise floor due to the decay of radioactive isotopes occurring in sea salt as the potassium-40 ($^{40}$K) is also present. Although the measurement of $^{40}$K was not among the goals of STRAW because the salinity is measured with high accuracy, the photons coincidence between the top and bottom PMTs of an sDOM  have been isolated revealing a pile up due to the  $^{40}$K. The level of  $^{40}$K measured is in excellence agreement with the salinity measured by ONC ($3.482\pm0.001$\% \cite{oceans2}), confirming a good understanding of the STRAW data \cite{Authors:2021htl}. \\

\noindent
The STRAW experience in the context of the large-scale ONC operation is an early demonstration that deep sea installation and operation of neutrino sensors has the potential today to achieve the scale of what would be needed for an extremely large-volume ocean-based neutrino telescope. 
STRAW was financed through the German Research Foundation (DFG) grant SFB\,1258 ``Neutrinos and Dark Matter in Astro- and Particle Physics'', cluster of excellence ``Origin and Structure of the Universe'', the University of Alberta and the Natural Science and Engineering Research Council of Canada (NSERC).\\

\begin{figure}[htp]
 \centering
 \centerline{
 \includegraphics[width=0.6\textwidth]{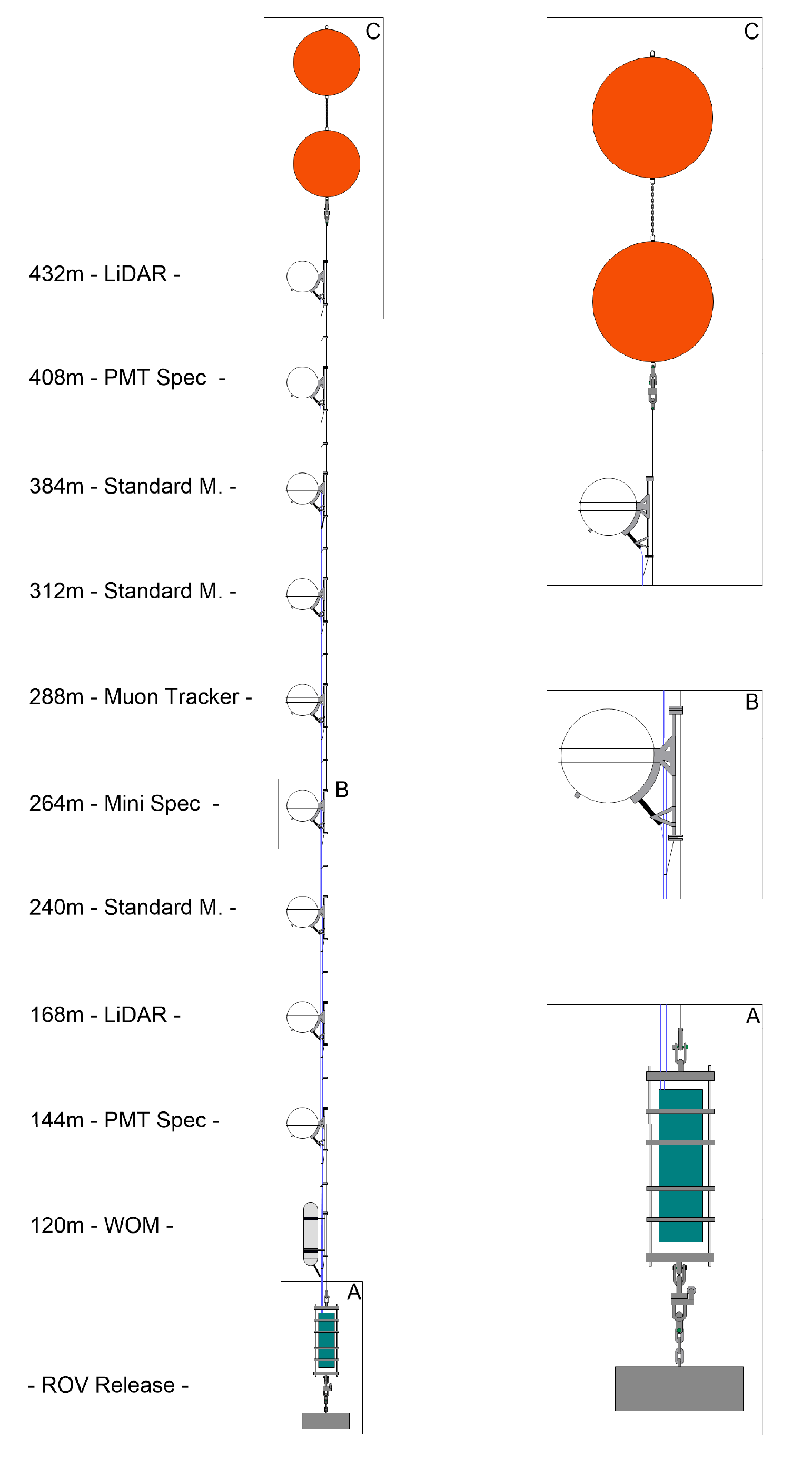}}
 \caption{{\small{\bf STRAW-b pathfinder: a bundle of 10 hybrid cables are equipped with modules. The modules are mechanically sustained by a backbone steel cable. As for STRAW, a \textit{mini Junction Box} connects the line to a main JB integrated in the NEPTUNE system.}}}
 \label{fig:Strawb_scheme}
\end{figure}

\noindent
{\bf The STRAW-b Pathfinder}\\
Having observed no significant challenges in the construction, installation, and operation of STRAW's two 145 meter long lines, and with the ambitious goal of building P-ONE with one-kilometer instrumented lines, we decided to challenge our capabilities by designing and building a 500-meter line named STRAW-b \cite{Rea:2021o3}. 
STRAW-b is equipped with ten modules all housed in spherical 13$''$ high-pressure resistant glass spheres. 
Three of the modules, named Standard Modules, monitor the environmental conditions (pressure, temperature, humidity) and host magnetic field sensors and accelerometers.
The other modules are two LiDARs (Light Detection And Ranging), a muon tracker, two  PMT-based spectrometers, a mini spectrometer, and a Wavelength shifting Optical Module (WOM) prototype built by the University of Mainz, see Fig. \ref{fig:Strawb_scheme}. Single cameras are also hosted in most of the modules to allow visual inspection.\\

\noindent
Completed in spring 2020 in Munich, the instrumented line was shipped to ONC where the Canadian part of the P-ONE collaboration took care of the final integration and tests. Deployment of STRAW-b occurred in Fall 2020 with less than optimal conditions due to both the pandemic and ocean conditions. Despite these additional challenges, the deployment with the anchor at the front of the line was successful. However, we lost connection with one of the standard modules right after the deployment. 
The most probable scenario is that one of the connectors failed, producing such a loss. The STRAW-b modules have been taking data continuously for several months and data are being integrated into the ONC database, Oceans 2.0
 (\url{https://data.oceannetworks.ca}). \\

\noindent
STRAW-b was financed through the German Research Foundation (DFG) grant SFB\,1258 ``Neutrinos and Dark Matter in Astro- and Particle Physics'', cluster of excellence ``Origin and Structure of the Universe'', and the University of Alberta.

\section{Start up P-ONE: the Prototype Line Project}
\noindent
Encouraged by the experience with STRAW and STRAW-b, since early 2021, we have started the design of a fully-instrumented P-ONE line \cite{Spannfellner:2021ro}. In addition to the knowledge gathered by P-ONE since 2017 and ONC over 15 years of experience, we integrated in the design phase lessons learned from IceCube, GVD, Antares, and KM3NeT.  
The P-ONE protoype line hardware will be optimized for minimal risks, maximal longevity, and scientific return. \\

\begin{figure}[htp]
    \includegraphics[width=0.8\textwidth]{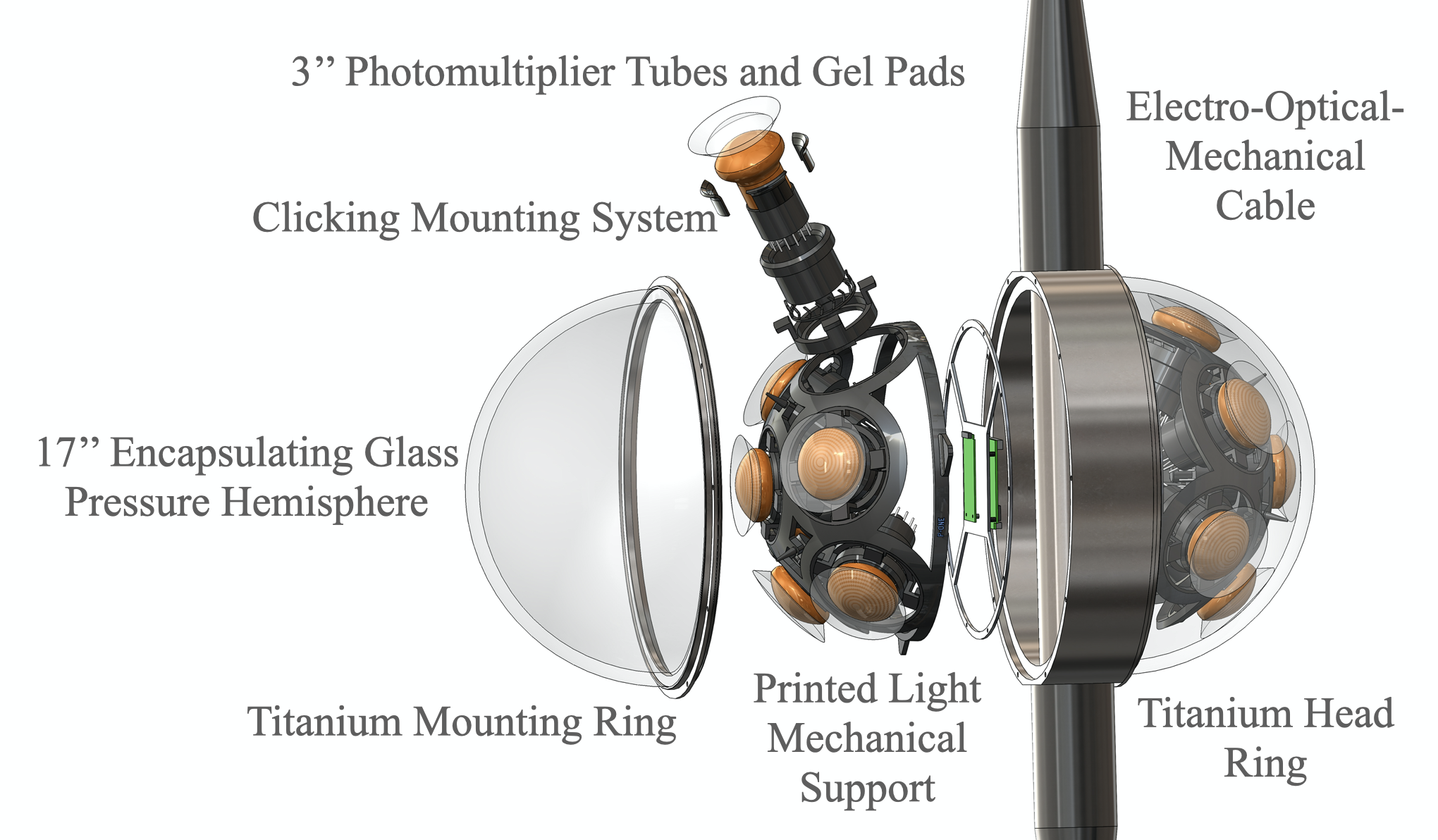}
    \caption[P-ONE]{{\small {\bf Fly’s eye anatomy of the P-ONE optical module. The integration of the active components of the telescope with the P-ONE cable will take place without the use of external connectors. The instrumented hemispheres will be assembled through titanium rings to an internal junction box. The details of these connections are still being finalised.  } 
}}
    \label{fig:OM}
\end{figure}

\noindent
A simplified integration between the backbone cable and the optical and calibration modules is guaranteed by 
a new connector-less cable design and sensor mounting that resembles the fly’s eye anatomy, Fig.\,\ref{fig:OM}.
Work is on-going on the finalization of the P-ONE optical module. As anticipated by KM3NeT work, to achieve the best possible performance within the high light-background environment of the deep ocean (O(10) kHz per PMT), a multi photo-multiplier tube (mPMT) configuration has been adopted. 
The photosensors in the P-ONE optical modules will be PMTs with a diameter between 3-3.5 inches. At the current development stage, candidate PMTs for the optical modules are being evaluated at calibration setups at the University of Alberta and the Technical University Munich. \\

\noindent
Experts from Michigan State University and the Technical University of Munich are developing the read-out electronics, time synchronization, and the data acquisition system based on state-of-the-art FPGA (Field Programmable Gate Array) and ML technologies to filter in-situ the extensive background, which is central to meet the boundary conditions of the ONC’s infrastructure.
For the first installation of up to 10 lines, bandwidth (4 Gbit/s) and power consumption (8 kW) available at the node will not be a limiting factor.
Looking ahead,  the challenge is to reduce bandwidth usage, keeping power consumption affordable. The time synchronization of the array will follow 
follow the White Rabbit protocol[, as in the future IceCube-Gen2 and the under construction KM3NeT telescopes \cite{Halliday:2021s1, CALVO2020162777}. A new calibration strategy developed by experts based a the Simon Fraser University, is also proposed for monitoring the floating geometry without using an additional acoustic triangulation system. This prototype phase aims to demonstrate the scalability, robustness, and maintainability of neutrino telescopes from the operation of ideally more than three prototype lines over several years.

\section{Realize and Operate P-ONE in the Global Context of Multimessenger Astronomy}
\begin{figure}[htp]
\centering
\includegraphics[width=0.8\textwidth]{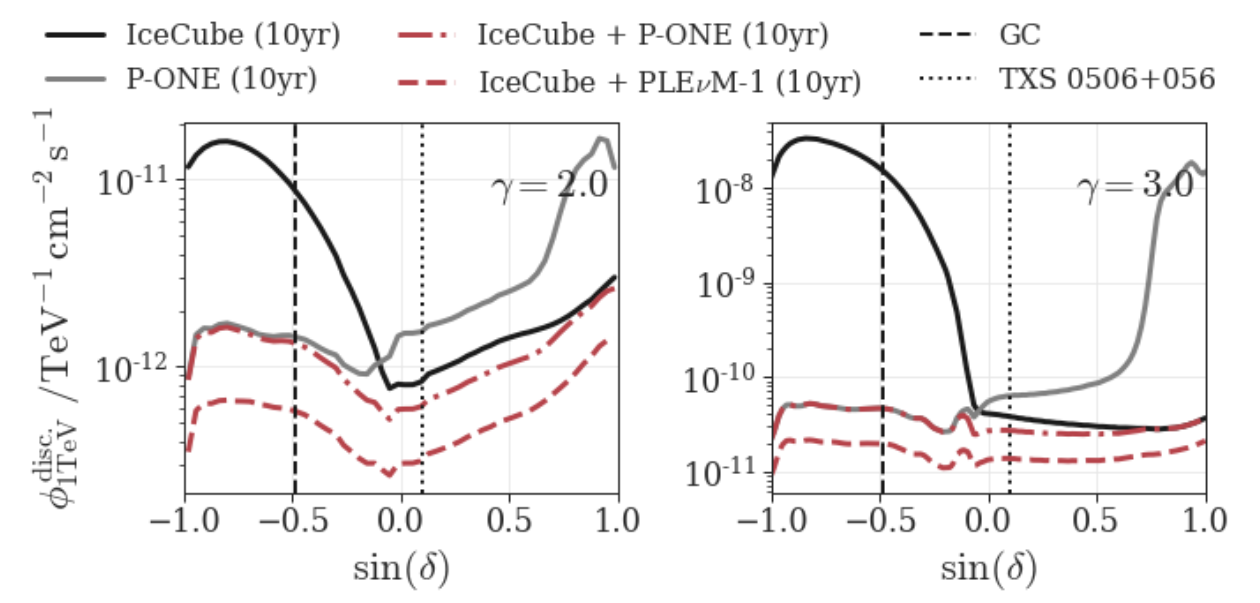}
\caption{{\small{\bf Evolution of the discovery potential for present (IceCube) and future neutrino telescopes (KM3NeT, GVD, P-ONE = PLE$\nu$M) and for two spectral indices: $\gamma=2.0$ (left) and $\gamma=3.0$ (right) \cite{Schumacher:2021Pv}. A synergetic operation among the neutrino telescopes will be decisive for the future of the field.}}
\label{Fig:comparison}}
\end{figure}

\noindent
IceCube has been foundational to neutrino astronomy. After 10 years of operation of IceCube, new radical steps, have to be take to advance the field beyond infancy. The field of neutrino astronomy must progress from the cubic-km scale of IceCube to the multi-cubic km era, and drastically expand coverage to open up a view of the entire sky, for the first time ever, through one network of telescopes. 
As studied in \cite{Schumacher:2021Pv}, a complementary operation among the multiple neutrino telescopes will be decisive and will boost the discovery potential in the Southern Hemisphere, an unexplorable land for a single telescope alone (Fig.\,\ref{Fig:comparison}), by up to three orders of magnitude. This will also address the need for expanded sky coverage for real-time astronomical observations with neutrinos all around the clock. 

\begin{figure}[htp]
\centering
\includegraphics[width=0.49\textwidth]{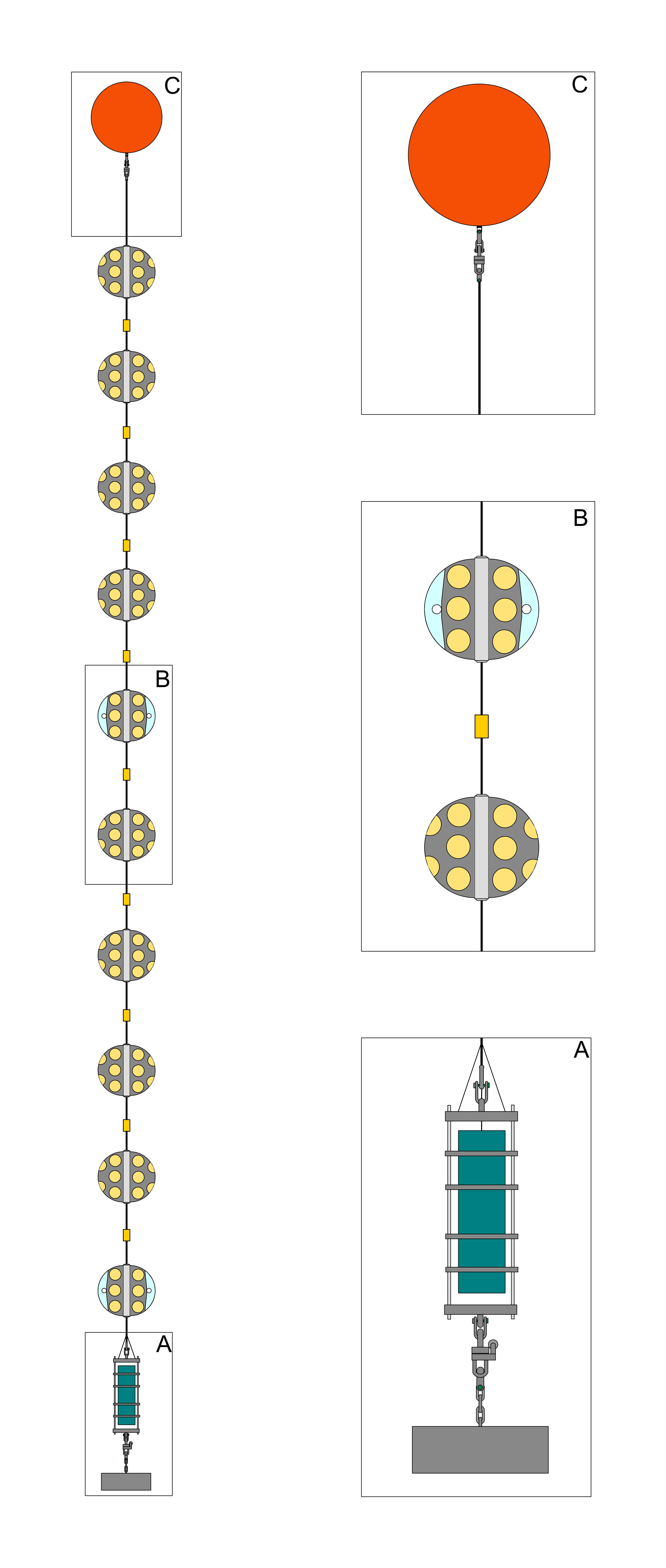}
\caption{{\small{\bf P-ONE prototype line under design. The line is one kilometer tall instrumented with 15 optical modules and 5 optical/calibration modules. The modules are engaged with the electrical-optical-mechanical cable through titanium rings which penetrates the module's head through a junction box that defends the vital elements of the cable from infiltration. The optical sensor hemispheres (eyes) are laterally mounted via a titanium flange which is glued to the glass hemisphere \cite{Spannfellner:2021ro}. }}}
\label{fig:line}
\end{figure}

\newpage
\bibliographystyle{unsrt}
\bibliography{bibliography}

\newpage
\section*{Full Authors List: P-ONE Collaboration}
%
%
\scriptsize

\noindent
Nicolai Bailly$^1$,
Jeannette Bedard$^1$,
Michael B\"ohmer$^2$,
Jeff Bosma$^1$,
Dirk Brussow$^1$,
Jonathan Cheng$^1$,
Ken Clark$^3$,
Beckey Croteau$^1$,
Matthias Danninger$^4$,
Tyce DeYoung$^6$,
Fabio De Leo$^1$,
Nathan Deis$^1$,
Matthew Ens$^4$,
Christopher Fink$^2$,
Rowan Fox$^1$,
Christian Fruck$^2$,
Andreas G\"artner$^5$,
Roman Gernh\"auser$^2$,
Darren Grant$^6$,
Dilraj Ghuman$^4$,
Christian Haack$^2$,
Rob Halliday$^6$,
Helen He$^1$,
Felix Henningsen$^7$,
Kilian Holzapfel$^2$,
Ryan Hotte$^1$,
Reyna Jenkyns$^1$,
Hamish Johnson$^4$,
Akanksha Katil$^5$,
Claudio Kopper$^6$,
Carsten B. Krauss$^5$,
Ian Kulin$^1$,
Klaus Leism\"uller$^2$,
Sally Leys$^8$,
Tony Lin$^1$,
Paul Macoun$^1$,
Thomas McElroy$^5$,
Stephan Meighen-Berger$^2$,
Jan Michel$^9$,
Roger Moore$^5$,
Mike Morley$^1$,
Laszlo Papp$^2$,
Benoit Pirenne$^1$,
Tom Qiu$^1$,
Mark Rankin$^1$,
Immacolata Carmen Rea$^2$,
Elisa Resconi$^2$,
Adrian Round$^1$,
Albert Ruskey$^1$,
Ryan Rutley$^1$,
Lisa Schumacher$^2$,
Christian Spannfellner$^2$,
Jakub Stacho$^4$,
Ross Timmerman$^1$,
Meghan Tomlin$^1$,
Matt Tradewell$^1$,
Michael Traxler$^{10}$,
Matt Uganecz$^1$,
Braeden Veenstra$^5$,
Seann Wagner$^1$,
Eva Laura Winter$^2$, 
Nathan Whitehorn$^6$, 
Juan Pablo Ya\~nez$^5$,
Yinsong Zheng$^1$\\
%
\\
\noindent
$^1$Ocean Networks Canada, University of Victoria, Victoria, British Columbia, Canada.\\
$^2$Department of Physics, Technical University of Munich, Garching, Germany.\\
$^3$Department of Physics, Engineering Physics and Astronomy, Queen's University, Kingston, Ontario, Canada.\\
$^4$Department of Physics, Simon Fraser University, Burnaby, British Columbia, Canada.\\
$^5$Department of Physics, University of Alberta, Edmonton, Alberta, Canada.\\
$^6$Department of Physics and Astronomy, Michigan State University, East Lansing, MI, USA.\\
$^7$Max-Planck-Institut f\"ur Physik, Munich, Germany.\\
$^8$Department of Biological Sciences, University of Alberta, Edmonton, Alberta, Canada.\\
$^9$Institut f\"ur Kernphysik, Goethe Universit\"at, Frankfurt, Germany.\\
$^{10}$Gesellschaft f\"ur Schwerionenforschung, Darmstadt, Germany.\\

\end{document}